\documentclass[12pt]{article}

\usepackage{graphicx}
\usepackage{fancyhdr}   
\usepackage{amsmath}    
\usepackage{comment}    
\usepackage{caption}
\usepackage{subcaption}
\usepackage{morefloats}

\usepackage{natbib}
\usepackage{lineno}
\modulolinenumbers[2]

\usepackage{setspace}
\usepackage{makeidx}  

\newcommand{\cov}{{\rm cov}}

\usepackage{setspace}

\usepackage{cite}
\usepackage{url}

\newcommand{\be}{\begin{eqnarray}}
\newcommand{\ee}{\end{eqnarray}}

\usepackage{amsfonts}
\usepackage{xspace}
\usepackage{epsfig}

\usepackage{setspace}
\date{}

\title{Is information a selectable trait?}
\author{Masoud Mirmomeni$^{1, 2,\dagger}$,
William Punch$^{1,2,\star}$, Christoph Adami$^{2,3,4,\ddagger}$}

\begin{document}

\maketitle

\begin{flushleft}
$^{\bf{1}}$ Department of Computer Science and Engineering, Michigan State University, East Lansing, Michigan, 48824 (USA).
\\
$^{\bf{2}}$ BEACON Center for the Study of Evolution in Action, Michigan State University, East Lansing, Michigan, 48824 (USA).
\\
$^{\bf{3}}$ Department of Microbiology and Molecular Genetics, Michigan State University, East Lansing, Michigan, 48824 (USA).
\\
$^{\bf{4}}$ Department of Physics and Astronomy, Michigan State University, East Lansing, Michigan, 48824 (USA).
\\
\vskip 1cm
$^{\dagger}$mirmomen@msu.edu\\
$^{\star}$ punch@cse.msu.edu\\
$^{\ddagger}$ adami@msu.edu (corresponding author)
\vskip 1cm
\end{flushleft}


\section*{Abstract}
There is little doubt in scientific circles that--counting from the origin of life towards today--evolution has led to an increase in the amount of information stored within the genomes of the biosphere. This trend of increasing information on average likely holds for every successful line of descent, but it is not clear whether this increase is due to a general law, or whether it is a secondary effect linked to an overall increase in fitness. Here, we use ``digital life" evolution experiments to study whether information is under selection if treated as an organismal trait, using the Price equation. By measuring both sides of the equation individually in an adapting population, the strength of selection on a trait appears as a ``gap" between the two terms of the right-hand-side of the Price equation. We find that information is strongly selected (as it encodes all fitness-producing traits) by comparing the strength of selection on information to a weakly selected trait (sequence length), as well as to a neutral marker. We observe that while strength of selection on arbitrary traits can vary during an experiment (including reversing sign), information is a selectable trait that must increase in a fixed environment.

\newpage
Whether or not complexity increases in biological evolution is a hotly debated topic. The controversy is, of course, intensified by public uneasiness with evolutionary concepts. However, even within firmly scientific circles the question is considered unsettled for a variety of reasons. While no agreed-upon definitions of complexity exist~\citep{Adami2002}, it is tempting to use information as a proxy for complexity, perhaps based on the vague notion that ``it takes information do produce something complex".  However, a precise definition of ``information" in the context of genomes is thought to be lacking. Furthermore, historical evidence of periods of extinction suggests that information may not always have been increasing. But even if it was stipulated that information increases, it is not clear whether this trend is active or passive, in other words, whether information is actively selected for in the evolutionary process, or whether it increases simply because it is correlated with another trait that is under selection.

While it is clear and obvious to every researcher in the field of evolutionary biology (and most often beyond) that information is ``that which resides in the genome", a precise quantification of information is difficult to obtain. Some equate information with alleles, some with genes. Some others argue that any genetic variation is information, while others again treat the entire genome of an organism as information. Others still argue that it is Fisher information that increases in natural selection~\citep{Frank2009}.
Within the last 15 years a consensus has emerged that the information content of biomolecular sequences (be they protein-coding, RNA coding, or non-coding, such as regulatory motifs) can be measured (or at least estimated) using techniques of Shannon information theory~\citep{Adami1998,AdamiCerf2000,Adamietal2000,Schneider2000,Szostak2003,TaftMattick2003,Carothersetal2004,Huangetal2004,Irizarryetal2005,Hazenetal2007,Leeetal2007,CarboneEngelen2009,Strelioffetal2010,Adami2012}. This information content has an intuitive meaning: it is the amount of information (in bits or any other suitable measure) that the holder has {\em about} the environment within which it thrives (we will give a mathematical definition below). As an organism uncovers (by the process of evolution) new ways to exploit and survive in that environment, its information about the environment increases. However, changes in the environment can lead to an abrupt (and sometimes catastrophic) decrease in information. Certainly the introduction of anti-viral or anti-bacterial drugs are information-decreasing to the pathogen, and the evolution of drug resistance can be viewed as the attempt of the pathogen to decipher the new environment and acquire the requisite information to exploit it again~\citep{GuptaAdami2014}. An alternative discussion of evolution as a process to discover information about the environment has been given by Frank~\citep{Frank2009,Frank2012b,Frank2013}, who uses a different definition of information. 

Measuring information changes along a (historical) evolutionary line of descent is not currently possible, as we would need to have access to extinct homologues of genetic sequences (but see~\citealt{Adami2012} for a phylogenetic approach). The evidence that information content increases in evolution therefore comes to us via computer simulations~\citep{Schneider2000} and instantiations of evolution via digital organisms~\citep{Adamietal2000}. In the latter study, a population of asexual digital organisms adapted in a single niche, and a first-order estimate of the information content of the sequences showed that information was steadily increasing as fitness increases. It is clear from this study and others that information has a fitness value~\citep{BergstromLachmann2004,Donaldsonetal2010}. But is this information actively selected for? To answer this question, we turn to an equation that promises to settle the issue: the Price equation~\citep{Price1970,Price1972,Price1995,Frank1995,Gardner2008,Frank2012a}. 

The Price equation is a statistical description of evolutionary change, and relates the changes in a trait variable over time--the left hand side (LHS)--to the covariance of fitness with the trait in question, as well as the expectation of a mutational change in said trait: both terms of the right hand side (RHS) of the equation (the extended Price equation of~\citealt{KerrGodfreySmith2008} has three terms).  Because of this identity, the Price equation is often described as a tautology~\citep{vanVeelen2005}, but it is so only if the LHS and the RHS are related mathematically. If they are related empirically, that is, each of the changes are determined by observing changes in an adapting population, the Price equation can reveal important aspects of the evolutionary dynamics at play.  For example, if fitness covaries with the trait, the first term is expected to be positive. Often times, the second term will compensate the first because most mutations on a (selected) trait are deleterious. Naturally, for traits that should be eliminated in order survive, the signs of both terms are reversed. Note that when a population is at mutation-selection balance (no mean change in the trait value), the two terms of the Price equation should cancel each other exactly. While our main focus in this article is to study whether information--treated as a trait value of an adapting organism--is under selection, to some extent we can also answer whether the Price equation is a useful tool to study adaptation. 

Treating information as a trait is certainly unconventional. It is clearly not on the same level as other observable (phenotypic) traits. Information is, however, a trait that is uniquely computable from the sequence (at least an estimate of information content is), and would be no different then using the GC content of a sequence~\citep{Hildebrandetal2010} or a particular transcription factor binding site sequence as a trait that covaries with fitness, or for example. But using information as a trait is different in another respect: Because {\em everything} that the organisms ``knows" about the environment is expressed in this information, it is the trait that should covary the most with fitness. In other words, no other variable should explain as much of the fitness of the organism as the information content. This suggests a very close relationship between information and fitness~\citep{Tayloretal2007,Polani2009,RivoireLeibler2011}, and we will discuss this issue below.

Clearly, using actual evolving populations (such as, for example, the Long-Term Evolution Experiment or LTEE, see~\citealt{Lenskietal1991,LenskiTravisano1994,Barricketal2009,Wiseretal2013}) to test the evolution of information would be preferable to all other methods. Unfortunately, as a precise application of Price's equation (see Materials and Methods) requires keeping track of all lineages generation after generation, this is essentially impossible. Our approach to study whether information is a selectable trait is as follows. We will use the digital life platform Avida~\citep{AdamiBrown1994,Adami1998,OfriaWilke2004,Adami2006} to perform evolution experiments that each focus on a different trait, allowing us unprecedented access to empirical data that is not usually available to the experimenter, such as the trait value for every single organism in the entire population, at every single generation. Experiments with `avidians' have proven to be sophisticated while remaining tractable (see, e.g.,~\citealt{Lenskietal1999,Wilkeetal2001,WilkeAdami2002,Lenskietal2003,Chowetal2004,Adami2006,Misevicetal2006,Ostrowskietal2007,Cluneetal2011,Covertetal2013,Goldsbyetal2014}). 

We study the information content (calculated from the sequence) as one trait, but for the purpose of comparison we will also use another trait (sequence length)  that we expect to be under weak selection, and that can change in either direction. We will compare these experiments to a control treatment where the trait we observe is constructed in such a manner that it cannot be under any positive or negative selection (a neutral marker). In this manner, we can probe the Price equation's ability to distinguish between traits that are under selection as opposed to those that are not, even while the empirical LHS and the RHS must be equal at least statistically, for each and every one of these cases. 

We find that information is a strongly selected trait as evidenced by a wide ``gap" between the two terms of the Price equation, while sequence length is weakly selected, and the neutral marker is not under selection. While in general a positive covariance between a trait and fitness does not imply a causal relationship between the two observables (either could cause the other, or they could be both caused by a third variable), it is difficult to escape the conclusion that information {\em causes} fitness. For morphological traits (and even metabolic ones) it is conceivable that high fitness could lead to changes in the trait (rather than the other way around as is more usual). In the present situation, it is hard to imagine how high fitness could cause increased information, as long as the population is well equilibrated. In turn, it is difficult to imagine that a third variable (such as a different trait) causes both high fitness (a possibility) as well as high information (not possible). Thus, we are left with the conclusion that, as intuition would have dictated to us all along, that the origin of an organism's high fitness in a particular environment is the genetic information it has accumulated about that environment. 

\section*{Methods}
\subsection*{\sc The Price Equation}
The Price equation~\citep{Price1970,Price1972,Price1995,Frank1995,Frank2012a} provides a statistical link between the mean change in a trait, and population-based averages of the covariance of trait and fitness, as well as the expectation of genetic trait changes (induced by mutations). 
This equation has been used to address a vast array of topics ranging from the evolution of group selection, kin selection, and cooperation~\citep{Wade1985,Queller1992,Frank1995,PageNowak2002,Henrich2004,FletcherZwick2006,Okasha2006} to ecology~\citep{Fox2006} and all the way to the selection of music loops~\citep{MacCallumetal2012}. The equation is usually written as follows:
\begin{equation}
\bar{w}\Delta \bar{z} = {\rm cov}(w,z) + E(w,\Delta z)\;. \label{Price}
\end{equation}
with \be
\cov(w,z)&=&\frac1{N_a}\sum_{i=1}^{n_a} n_i(w_iz_i-\bar w \bar z)\; , \\
E(w,\Delta z)&=&\frac1{N_a}\sum_{i=1}^{n_a} n_i w_i(z^\prime_i-z_i)\;.
 \ee
Here,  $w_i = \frac{n^{\prime}_i}{n_i}$ and $z_i$ are the fitness of group $i$ (typically, a genotype) and the trait, respectively, in the ancestral population, 
$n_i$ and $n^{\prime}_i$ are the total number of organisms of that type in two consecutive populations (ancestors and descendants), and 
$\Delta z_i = z^{\prime}_i - z_i$ is the amount of change in character in group $i$ between descendant and ancestral (for example by a mutation). 
The trait could in principle be anything (eye color, resource affinity, mating behavior, etc.) and could be positively or negatively associated with fitness (or it could be entirely neutral).  The average fitness in the ancestral population is given by 
\be
\bar{w}=\frac1{N_a}{\sum_{j=1}^{n_a}{n_j w_j}}\;,
\ee while 
$\Delta \bar{z} = \bar z^{\prime} - \bar z$ is the average amount of change in the trait in two consecutive generations,  given by:
\begin{equation}
\bar z=\frac1{N_a}{\sum_{i=1}^{n_a}{n_i z_i}}\;,\;\;\;\;
\bar z^{\prime}=\frac1{N_d}{\sum_{i=1}^{n_d}{n^{\prime}_i z^{\prime}_i}}\;.
\label{eq:01b}
\end{equation}
In the above equations, $n_a$ refers to the number of ancestral types, and $N_a=\sum_{i=1}^{n_a}n_i$ is the ancestral population size (replace `a' with `d' for the descendants).  

Recently, a generalized form of this equation was derived~\citep{KerrGodfreySmith2008} in which immigration and recombination can be considered as genetic sources (in the equation above, the only source of a genotype was from its ancestral type or a mutation). For simplicity, in this work we use the original Price equation as recombination and immigration are absent for the populations we study.

\subsection*{\sc Avida: a Virtual Environment for Digital Organisms}

Avida creates a virtual environment within standard desktop computers in which self-replicating computer programs (called ``digital organisms" or ``avidians") compete for resources, mutate, and evolve~\citep{AdamiBrown1994,Adami1998,WilkeAdami2002,OfriaWilke2004,Adami2006}. Avidians can inhabit structured environments (such as would be encountered by bacteria on a Petri dish), or well-mixed environments exemplified by chemostats. As the computer programs must replicate (make copies of the genomic information) to survive, evolutionary `runs' are experiments much more than simulations, as the information must be physically copied (line by line) from location to location in computer memory, and evolution proceeds via point mutations (and possibly other types such as insertions and deletions) that occur during this process of copying. A point mutation occurs when (with a user-specified probability) a random instruction is written into the daughter genome in place of the instruction that was read. 

To be successful, an avidian's growth rate (related to the number of offspring it generates per unit time, the `update') must exceed the growth rate of other organisms present in the population, and this growth rate depends on the efficiency with which the organism is able to copy its genome, as well as how well it is able to exploit its environment. Avidians obtain the necessary energy to replicate from performing {\em computations} (which we view as the digital analogue of an organism's metabolism), which allows it to collect resources that can be converted into energy. These computations are mathematical operations performed on random binary numbers that the environment provides, and only a specific sequence of instructions (or variants thereof) trigger the resource and concomitant energy. An increased energy buys the organism a ``faster" CPU (central processing unit), which in turn increases the growth rate. The resources are  available at equal concentrations to every organism, approximating a well-stirred chemostat. In the experiments reported here, the resources are unlimited (meaning that they are not depleted by use), which creates a single-niche environment for organisms without negative frequency-dependent selection. In the present experiments, we reward nine different bit-wise logic operations (such as AND, OR, NOR, etc.) an environment commonly referred to as the ``logic-9" environment. 

In the treatment using information as a trait (as well as for the neutral trait), we limited our population size to 400, which is considerably smaller than the more ``standard" population sizes that are an order of magnitude larger. We used smaller population sizes in order to render the computational problem of tracking each individual's lineage in the population tractable. While smaller population sizes create the opportunity for random genetic drift, they are large enough so that we can observe the standard evolutionary dynamics in asexual populations~\citep{Elenaetal2007}.

Time in Avida is measured using two different measures. One `update' has gone by if every organism in the population has executed a fixed number of instructions on average (here set to 30, the default). Because there are 400 organisms in the population, an update has elapsed if $400\times 30$ instructions have been executed. Because it takes organisms hundreds of instructions to create an offspring (this depends on the sequence length, the efficiency of the copying algorithm, and the number of computational tasks executed) many updates can go by before 400 births have taken place, which marks the passing of one generation. Because the Price equation considers subsequent generations when connecting descendant to ancestral populations, we use generations as the unit in time. A complication arises because generations are overlapping in Avida (as they are in most biological populations) so that the same genotype often straddles one (or even more) generations, while the Price equation assumes that, generation by generation, populations are replaced in their entirety.  To account for this, when calculating the average change in a character in two consecutive generations, we change the parent identifier (ID) of those descendant entities that straddle both populations (descendant and ancestral) from their original parent ID to their own ID. 

The experiments reported in this paper were performed using version 2.12 of the Avida software which is freely available from avida.devosoft.org. The sequence length of the ancestral organism (called `organism.heads.100') that we used to seed every reported run is 100 instructions. This simple organism can self-replicate but is not able to accomplish any logical computations. For more details about Avida, including the default set of instructions and the definition of the environment, refer to~\citep{OfriaWilke2004}, and to the web resource above. Experiments ran under the Linux operating system on an Intel10 1536 cluster of 192 dual core Intel 2200+ processors at Michigan State University's High Performance Computing Centre.

\subsection*{\sc Fitness and Information Content of Digital Organisms}

The fitness of a genotype $w$ in a particular environment is defined by a typical representative's {\em expected} rate of offspring production.
This definition makes it easy to apply the Price equation to a population of Avidians.
Fitness can be {\em estimated} by the ratio of necessary energy for an organism to execute the instructions in its genome and the required number of instructions to produce an offspring~\citep{Adami1998,Elenaetal2007}, but the realized fitness depends on the organism's interaction with others and the vagaries of chance. The mandatory energy for reproduction is provided to a digital organism in the form of discrete quanta, called ``Single-Instruction Processing" units, or SIPs~\citep{Lenskietal2003,Adami2006,Elenaetal2007}. The rate at which an organism acquires energy depends on its genome length and a coefficient which is related to the interactions between the organism's computational metabolism and its environment. In other words, an organism gets more SIPs if it accomplishes specific computational functions in an environment that rewards those functions. As a consequence, it is easy to compare the relative rates of reproduction of any two given organisms. If an organism has $w$ twice as large as another organism, then it is expected that the former organism produces offspring at twice the rate of the latter. Moreover, organisms that cannot produce any offspring have zero fitness. The absolute fitness of the ancestral organism (`organism.heads.100') is 0.2494, which we use to normalize the fitness of the evolved populations relative to this ancestor. Note that this fitness is calculated by allowing the organism to grow in a single virtual dish, and so is an estimator of long-term success. During an experiment, fitness is not calculated for any organisms: either they survive and leave offspring or they do not. When calculating the information content of single sequences as described below, we have to estimate the fitness of all possible one-mutants of that sequence, which is again obtained via virtual growth-assays in isolation. 

In general, information is defined by Shannon as the shared entropy between two ensembles~\citep{Shannon1948,CoverThomas2006} and therefore cannot be determined for a single sequence. To circumvent this, we use a form of information that is conditional on a fixed environment (the environment that the organism actually encounters). For an organism with a fixed sequence length $\ell$, we can calculate the maximal entropy (uncertainty) for a hypothetical infinite population: this would be $H({\rm max})=\ell$ (if we agree to take logarithms to the base $D$, where $D$ is the number of possible instructions, that is, the alphabet size) because the likelihood to find any particular sequence in such an ensemble is $p=1/D^\ell$. In the present experiments, $D=26$.

This maximal entropy only describes {\em unselected} (that is, random) sequences. Those that accomplish a function (and thus replicate) have a very different entropy $H_{\rm actual}$, which depends on the relative concentration of each of the existing types in the population.
This entropy depends on the environment (thus it is a conditional entropy). If we associate the genetic sequence of a particular organism with the states of a random variable $A$, and describe the environment as the specific state $E=e$ of an environment random variable, we can write the information content of organism $A$ as~\citep{Adami1998,AdamiCerf2000}
\be
I(A)=H(A)-H(A|E=e)\;, \label{comp}
\ee
where we have followed the convention that unconditional entropies are maximal, i.e., $H(A)=H_{\rm max}=\ell)$ and
\be
H(A|E=e)=-\sum_{a} p(a|e)\log_D p(a|e)\;, \label{cond}
\ee
where $p(a|e)$ is the probability to find the self-replicating type $A=a$ given that the environment is in state $E=e$. 
The actual entropy $H(A|e)$ in Eq.~(\ref{cond}) is in principle defined via a sum over an infinite ensemble of sequences (all sequences $a$ that, given $e$, are functional). We will simulate such infinite ensembles artificially via an exhaustive mutational process (described below), because sampling the frequencies of organisms in an actual (small) evolving population cannot approach the true entropy (\ref{cond})~\citep{Huangetal2004}. 

Generally speaking, the entropy of a polymer can be written in terms of the sum of the entropy of the monomers (at each site), the pairwise entropies (shared entropy between sites), shared entropy between triples of sites (and so on), using a formula due to Fano~(\citeyear{Fano1961}). Writing the random variable $A$ in terms of the monomeric random variables $A^{(i)}$
\be
A=A^{(1)}\times A^{(2)}\times ... \times A^{(\ell)}
\ee 
the entropy $H(A|e)$ can be written as~\citep{Fano1961}
\be
H(A|e)=\sum_{i=1}^\ell H(A^{(i)}|e)-\sum_{i<j}^\ell H(A^{(i)}:A^{(j)}|e)+\sum_{i<j<k}^\ell H(A^{(i)}:A^{(j)}:A^{(k)}|e)-\cdots  \label{full}
\ee
In (\ref{full}), the sum goes over alternating signs of correlation entropies, culminating with a term $(-1)^{\ell-1}H(A^{(1)}:A^{(2)}:\cdots :A^{(\ell)}|e)$.  Here, we neglect all higher-order correlation terms and assume that the entropy $H(A|e)$ can be approximated by the sum over the per-site entropies $\sum_{i=1}^\ell H(A^{(i)}|e)$. In that case, Eq.~(\ref{comp}) can be written simply as 
\be
I(A)=\ell-\sum_{i=1}^\ell H(A^{(i)}|e)\;. \label{info}
\ee
The entropies $H(A^{(i)}|e)$, in turn, can be obtained by creating all single mutants of organism $A$, and measuring their fitness in isolation. We count all single mutations at site $i$ that are either neutral (or in rare cases beneficial) to obtain $\nu_i$ and estimate the entropy of site $i$ as $\log_D\nu_i$, as in~\citep{Adamietal2000}.  By taking logarithms to the base $D$, we normalize the entropy in such a way that a single monomer (a single instruction) has maximal entropy 1 `mer'~\citep{Adami2004}. Naturally, the maximal entropy of an $\ell$-mer then is $\ell$ mers, which is also the maximal information (for $D=26$, one mer equals approximately 4.7 bits). 

Neglecting the sub-leading terms $\sum_{i<j}^\ell H(A^{(i)}:A^{(j)}|e)$ can have significant repercussions (see, e.g.,~\citealt{GuptaAdami2014}), but we do not expect them to affect the conclusions we reach about selection even though they will affect the quantitative estimate of the information content.

\section*{Results}
We test whether three different traits (information, sequence length, and a neutral marker) are evolutionarily selected, by measuring the mean rate of change of the trait (the LHS) of the Price equation, and in particular the two terms on the RHS of the Price equation.  
\subsection*{\sc Information as a Selectable Trait }
We set up an experiment to record fitness and information content (as the trait)
for every organism in the population at every generation as described in Methods, for 14,000 generations. The sequence length is fixed to remain at 100 instructions, while the point mutation rate is set to $\mu=0.015$ so that on average each genome receives 1.5 mutations per generation. Insertion and deletion mutations are turned off. We chose the relatively high mutation rate so that significant evolutionary changes can be obtained with a comparably short experiment of 14,000 generations (still requiring $5.6\times 10^6$ trait evaluations, each of which requires 2,600 fitness evaluations: one for each substitution at each of the 100 sites, for a total of over 10 billion fitness evaluations per run). 

\begin{figure}[!htbp]
\begin{center}
                       \includegraphics[width=0.65\textwidth]{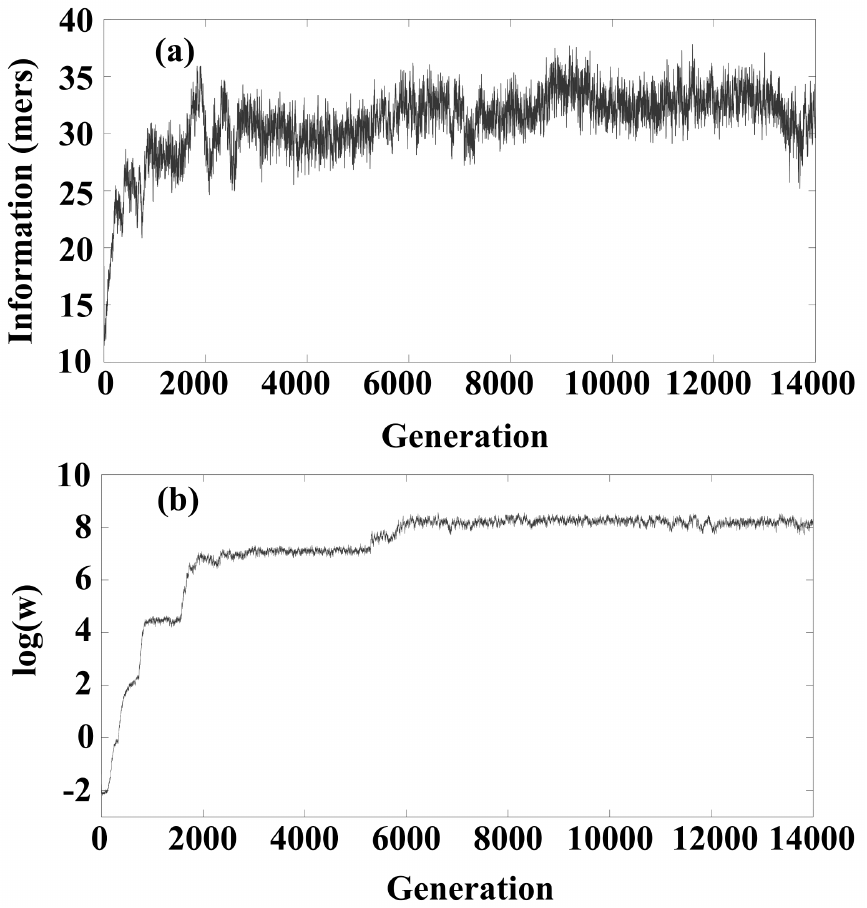}
                    \caption{\small{Average fitness and information content in a population of avidians through evolution. (a): average population information content. (b): Average absolute fitness.}}
        \label{fig:01}
\end{center}
\end{figure}

We show in Figure~\ref{fig:01} the mean fitness and average information content over 14,000 generations. Fitness shows the usual stair-step increase (note that we plot the logarithm of fitness). The 100 instruction-long ancestral sequence consists of a self-replicating `gene' that spans 15 instructions, and 85 inert (meaning, non-functional) {\tt nop-C} instructions. Because the fitness of the organism is unlikely to change if the inert instruction is replaced by any of the other $D$ instructions, each contributes $\log \nu=\log D=1$ to the conditional entropy $H(A|e)$, which subtracts from the first term in (\ref{info}). Thus, the information content begins at about 15 (it is slightly lower because the replication gene has some redundancy in it). Information content increases monotonically at first as more and more of the 85 ``empty" instructions are filled with information, but later we can see periods where the measured information decreases. 

We attribute these decreases to our use of the rough estimate (\ref{info}) as a proxy for the full information content that uses Eq.~(\ref{full}). While this approximation introduces noise, we are forced to do so because even including the first correction term $\sum_{i<j}^\ell H(A^{(i)}:A^{(j)}|e)$ would make the computation almost intractable. Fluctuations in average information content are likely due to the high mutation rate. 

\begin{figure}[!htbp]
\begin{center}
                \includegraphics[width=0.75\textwidth]{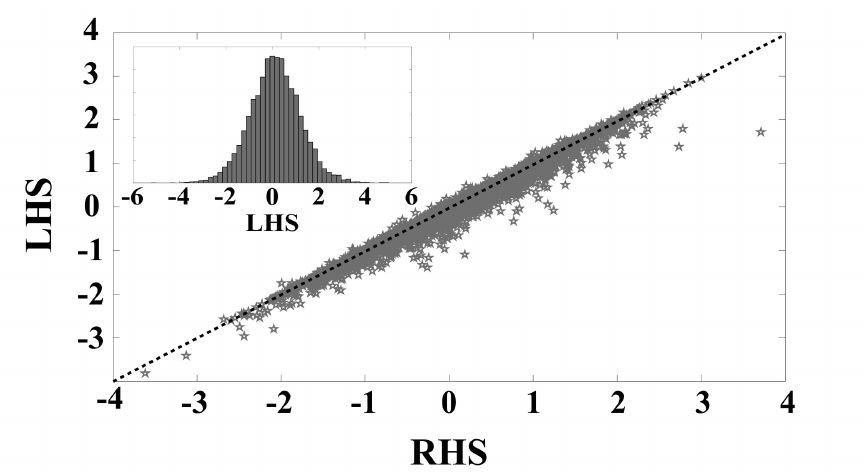}     
\caption{\small{Plot of the LHS vs.\ the RHS of the Price equation (\ref{Price}) with information as trait. The dashed line is the identity line, and the inset show the distribution of the LHS of the Price equation), which is approximately normal with a mean shifted to the right of zero.}}        
        \label{fig02}
\end{center}
\end{figure}

Using the average {\em change} in information in subsequent generations $\Delta\bar z$ and a calculation of the two terms on the RHS of the Price equation (\ref{Price}) directly from the population, we first test whether the Price equation is obeyed. Fig.~\ref{fig02} confirms that this is the case: we plot the LHS vs.\ the RHS, and note that they cluster tightly around the identity line. We also note that if the Price equation were a mere tautology, we should obtain a perfectly straight line. Instead, we observe that the data cluster slightly below the identity line, which is likely an artifact caused by the small population size (the bias is absent in the second treatment where the population size is much larger).  In the inset of Fig.~\ref{fig02} we show that the distribution of the LHS is slightly skewed to the right, implying that there are more transitions towards higher information than there are towards lower information. This observation simply corroborates that, on average, information is increasing. 

We now turn to analyzing the LHS and the RHS of Eq.~(\ref{Price}) separately. The first term, which also appears in the Price equation without mutations, quantifies the covariance of fitness with the trait. Indeed, in the absence of mutations, it is this term that drives the mean change of trait (the LHS) positive or negative (depending on whether the covariance is positive or negative). It thus reflects the force of purifying selection: it vanishes once all variance in the population has disappeared. Note that if fitness was used as a trait, the covariance would simply be the variance of fitness (the covariance of fitness with itself), and the Price equation would become Fisher's Fundamental Theorem as has been noted repeatedly, not the least by Price himself~\citep{Price1972b}. 

The second term of the RHS takes mutations into account explicitly, as without mutations $\Delta z=0$ (the trait could not change from generation to generation, only the frequency of traits could). It is this term that drives adaptation: single rare mutations can give rise to large positive $\Delta z$. However, in most cases, $\Delta z$ will be negative, as mutations disrupt the trait. Thus, the distribution of $\Delta z$ reflects the distribution of beneficial, neutral, and deleterious mutations. And because most mutations are detrimental, the second term of the RHS is mostly negative even though the trajectory of increasing $\bar z$ is driven by the rare beneficial mutations. 

\begin{figure}[!htbp]
\label{leftright}
\begin{center}
                \includegraphics[width=0.65\textwidth]{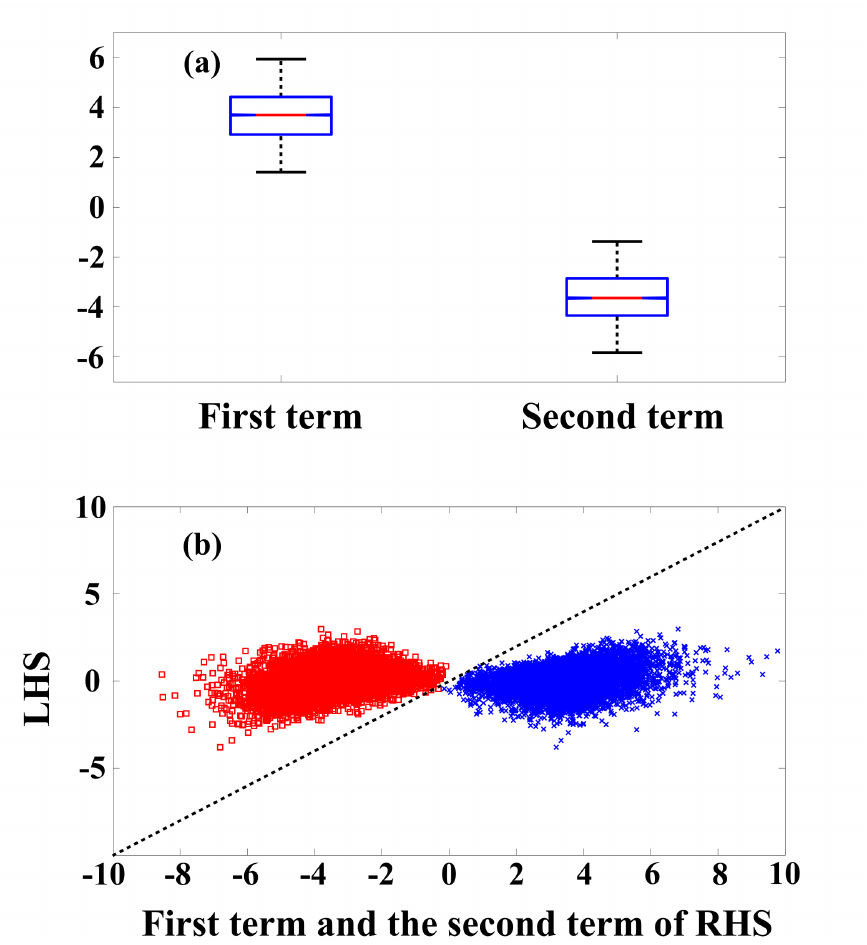}
\caption{\small{Opposing contributions of first and second term of RHS of Price equation. (a): Size of first and second term, averaged over the 14,000 generations of a typical experiment. The box represents the first and third quartile while the red line represent the median, and the whiskers encompass 99\% of the data. Notches (blue) indicate 95 percent confidence interval of the median. 
(b) LHS of Price equation vs.\ first (blue dots) and second (red dots) term of the RHS separately (we show a random sample of 7,000 out of 14,000 points), showing that the counterbalancing of two terms is necessary to satisfy the Price equation. }}             
        \label{fig03}
\end{center}
\end{figure}

In Fig.~\ref{fig03}a we show the size of first and second term of the RHS of the Price equation separately, and see that indeed the first term is positive on average, and the second term is negative on average. It is selection that drives these terms apart (as we will also show explicitly in the second treatment and the control): the first term shows strong covariance of information with fitness, and the second term shows that this positive association is balanced by mutations that tend to destroy the trait. In Fig.~\ref{fig03}b we plot the LHS against the first and second term separately, and show that neither lies on the identity line by themselves. The clear separation between the first term and the second term--which we term the ``Price equation gap"--is an indication of the strength of selection operating on the trait. We would like to stress that using the first term alone to assess strength of selection on a trait is not sufficient, as {\em any} trait is likely to change as the population approaches a pure (meaning homogeneous) state. The Price equation gap, on the other hand, is a strong indicator as it reflects the forces of selection in mutation-selection balance.

\subsection*{\sc Weak Selection on Sequence Length}
In order to be able to appreciate the nature of selection on information, we also study selection on a more ``classical" trait (the sequence length) using the Price equation.  In the first treatment, sequence length was kept constant so as not to confound selection on length with selection on information. For this control treatment, we allow sequence length to change using insert and delete mutations at a rate of 0.01 and 0.001 per genome per generation respectively, along with the copy-mutations at a rate of $\mu=0.00175$ per instruction copied, and a constant population size of 1,400 individuals. Note that the ratio of insert to delete mutations is biased towards an increase in sequence length if the sequence was drifting neutrally.  

\begin{figure}[!htbp]
\begin{center}
                 \includegraphics[width=0.65\textwidth]{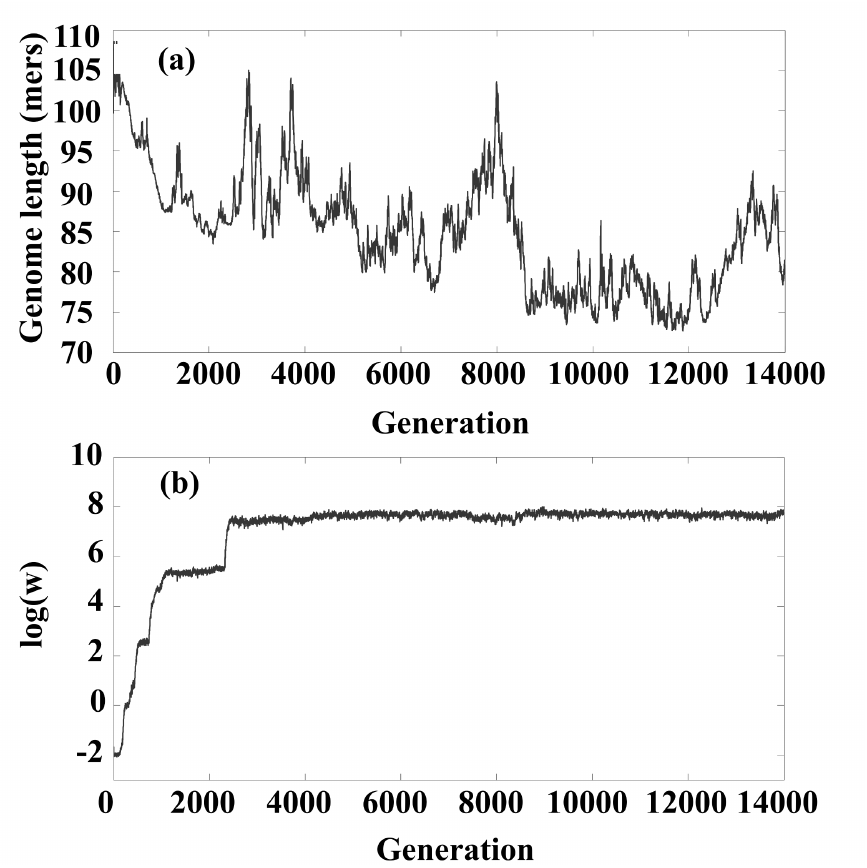}
             \caption{\small{Population-averaged sequence length and fitness. (a): Mean sequence length as a function of time (in generations). (b): Mean absolute fitness as a function of time (in generations).}}
        \label{fig04}
\end{center}
\end{figure}

We seed the experiment with the same ancestor as the previous experiment, but expect sequence length changes to occur that are adaptive. While extra instructions are not directly detrimental (or beneficial) to an individual's fitness because each organism obtains SIPs in proportion to sequence length~\citep{OfriaWilke2004}, there is a weak selection pressure to reduce sequence length so as to reduce mutational load~\citep{Lenskietal1999}. There is also a pressure to increase sequence length to accommodate new beneficial genes, but only if the current sequence length is insufficient to accommodate the new information. Thus, we expect to see selection in both directions, and indeed we witness first a trend towards reduced sequence length (as this allows the population to reduce the mutational load) and later intermittent trends towards higher sequence length (see Fig.~\ref{fig04}a). Fitness is steadily increasing during this period of changing sequence length, as evidenced by Fig.~\ref{fig04}b.

\begin{figure}[!htbp]
\begin{center}
                     \centering
                \includegraphics[width=0.75\textwidth]{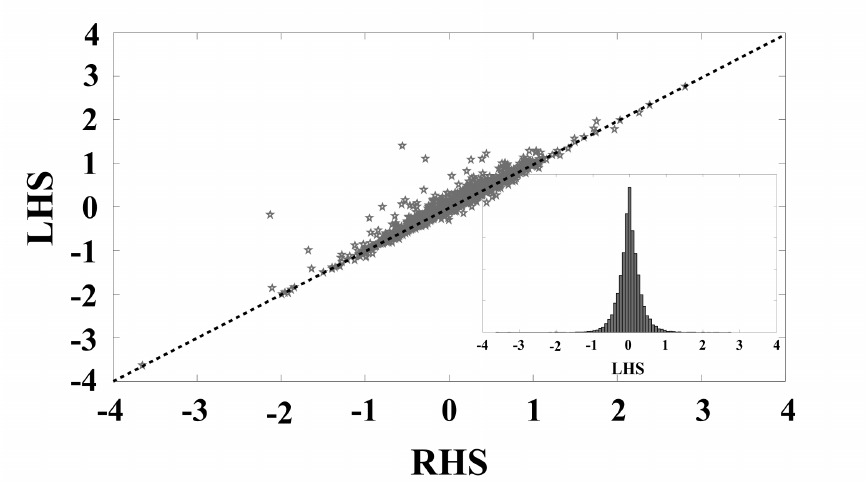}
                \caption{LHS vs.\ RHS of the Price equation for sequence length as trait. Each dot represents a (LHS,RHS) pair along the evolutionary trajectory (we show 7,000 randomly selected points out of 14,000). The dashed black line represents the identity line. The inset shows the distribution of trait changes (the LHS).}
                \label{fig05:A}
               \end{center}
\end{figure}

We record all individual genomes (along with the trait under consideration, the sequence length), and their fitness for 14,000 generations, and calculate mean changes in trait every generation (for the LHS), as well as the covariance and the expectation of trait change (for the two terms of the RHS).
We first verify that the Price equation is obeyed by plotting LHS vs.\ RHS (see Fig.~\ref{fig05:A}). Note that the distribution of length changes (shown in the inset of Fig.~\ref{fig05:A}) is slightly biased towards decreases. 

\begin{figure}[!htbp]
\begin{center}
                \includegraphics[width=0.65\textwidth]{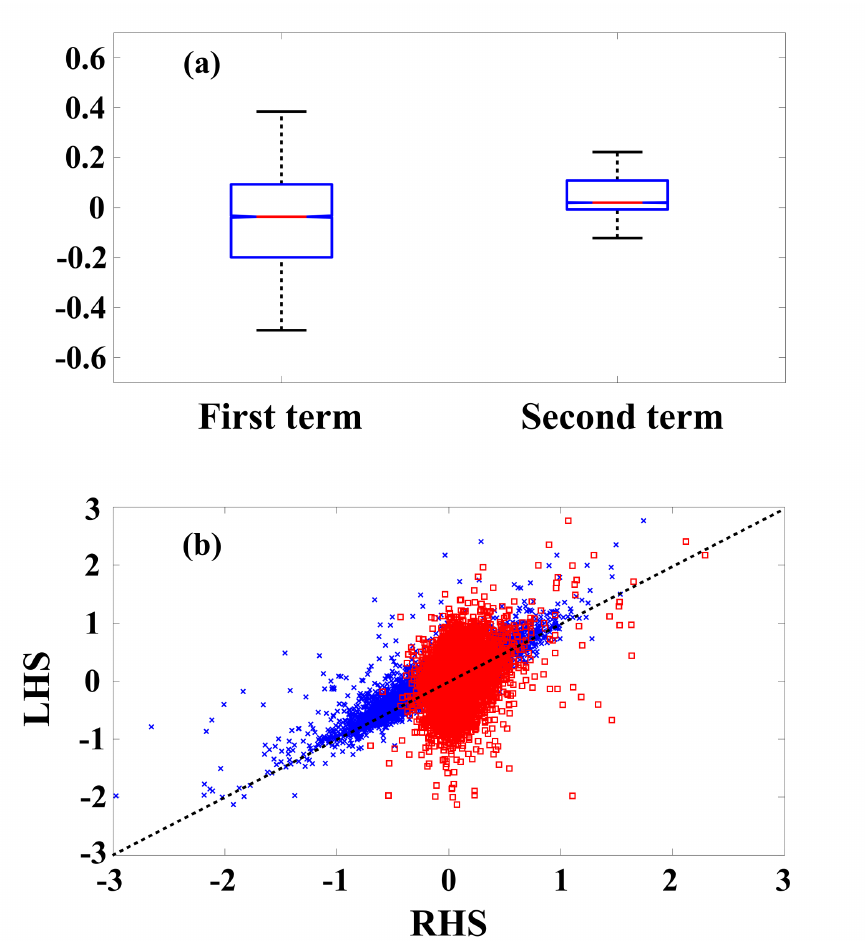}
\caption{\small{Contributions to the RHS of the Price equation for sequence length trait. (a): A comparison of the size of the first and second term shows that they are  significantly different from zero and from each other.  The box again represents the first and third quartile while the red line represent the median, and the whiskers encompass 99\% of the data. Notches (blue) indicate 95 percent confidence interval of the median. Note the small range of the trait. (b): Scatter plot of LHS vs.\ first (blue points) and second term (red points) of Price equation individually shows that the distributions overlap (random sample of 7,000 out of 14,000 points shown).}}           
        \label{fig6}
\end{center}
\end{figure}

Next, we study the Price equation gap by plotting in Fig.~\ref{fig6}a the size of the first term and the second term of the RHS individually as before. Both the first and second term medians are significantly different from zero (the notches in the box plot are 95\% confidence intervals) and from each other, but the effect is small. The overall Price equation gap (averaged over the entire evolutionary history) is small. 
Plotting the LHS vs.\ the first and second terms of the RHS independently (as in Fig.~\ref{fig03}b) shows that the two terms are not well separated, and instead overlap significantly. The mutation term (red dots in Fig.~\ref{fig6}b) seems to skew towards positive changes, which could be attributed to the biased insert mutation rate rather than to a selective pressure favoring increases in the trait. 
We conclude that sequence length is under selection, but only weakly so (the first term is weakly negative, and the second term is weakly positive) when averaged over the entire trajectory. The overall signal of selection strength can be weakened when selection acts in opposing directions at different times, and we should expect to see stronger signals if we would focus on those parts of the trajectory where sequence length is exclusively under positive or negative selection.

\subsection*{\sc Price Equation for a Neutral Marker}
As a control for the two previous treatments we test the Price equation on a marker that is by construction completely neutral.  
In this experiment, we study as a trait the difference in frequency between two specific neutral ``alleles", the instructions {\tt  nop-A} and {\tt nop-B}. These ``no operation" instructions have a role in the function of a typical avidian genome as they can specify the register that is used in internal calculations of the CPU, or else they can mark areas of code for recognition (pattern-based addressing). However, the frequency of this instruction is not at all under selection, and adding or deleting this instruction in code that is not specifically involved in computations is entirely neutral. In particular, the {\em difference} in frequency between those two instructions is expected to be unselected even if the absolute value of either was. Note that we did not use the third no-instruction operation {\tt nop-C} as a neutral marker as the ancestral genome is filled with 85 of those instructions, so reducing this number will be under strong selection initially. Because we collected the data for the neutral trait at the same time as we collected the information trait, we can use the same experiment. 

\begin{figure}[!htbp]
                \centering
                \includegraphics[width=0.75\textwidth]{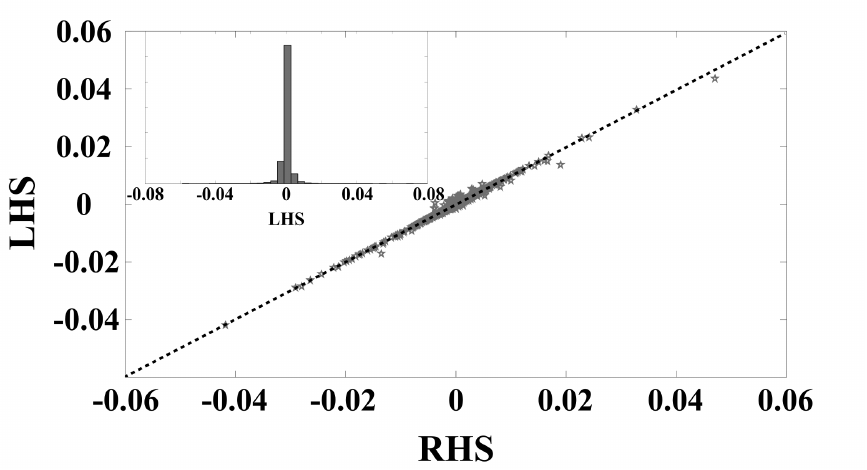}                  
        \caption{LHS vs.\ RHS of the Price equation for a neutral trait. The Price equation is strictly obeyed, but the changes in trait value (LHS) as well as the covariance and expectation of trait change are much smaller than in the previous treatments (note the scale on the axes). The distribution of trait changes (see inset) is symmetric and strongly peaked at zero.}
           \label{fig7}
\end{figure}

As expected, the trait does not change significantly during evolution, as a plot of the LHS vs. the RHS of the Price equation for this trait reveals (see Fig.~\ref{fig7}). 
\begin{figure}
                \centering
                \includegraphics[width=0.65\textwidth]{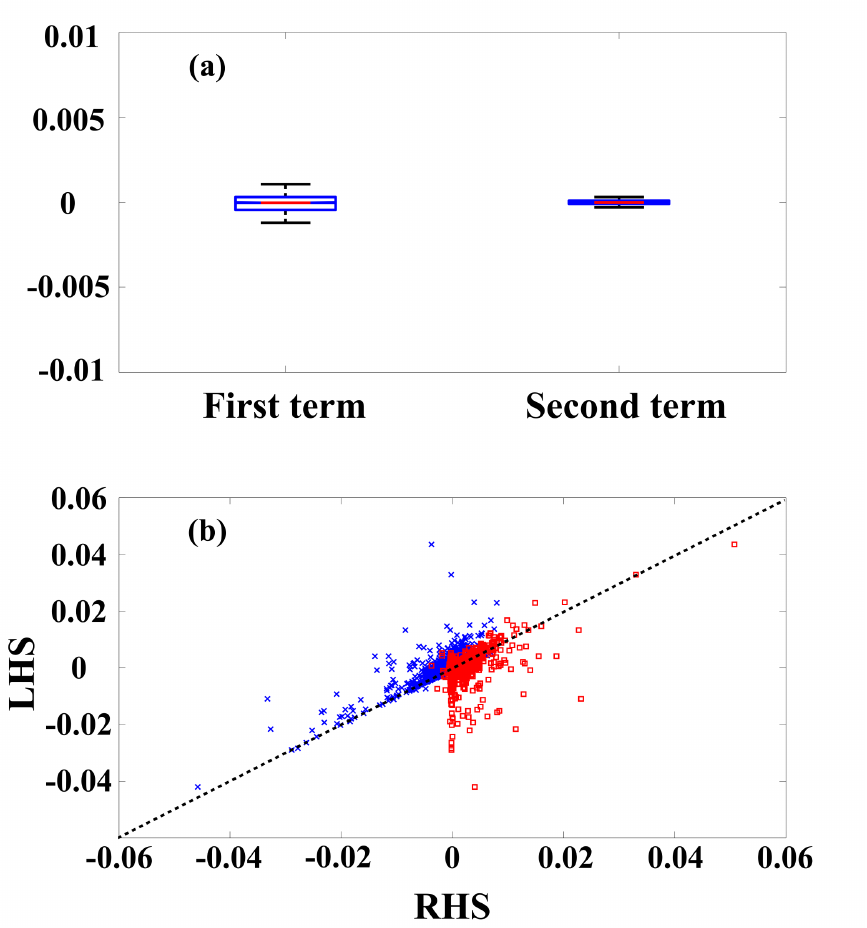}
\caption{\small{Contributions to the RHS of the Price equation for a neutral marker trait. (a): A comparison of the size of the first and second term shows that they are not significantly different from zero and from each other. Boxplot as in Fig.~\ref{fig6}b}. (b): Scatter plot of LHS against first term of RHS (blue) and second term of RHS (red) of the Price equation for the neutral trait.}          
        \label{fig8}
\end{figure}
We can also study the Price equation gap, and find it to be non-existent (see Fig.~\ref{fig8}a), but not just on average, but for the entire experiment. There is little variation in the LHS or RHS, and no difference between the two terms of the RHS, and with zero. Plotting the two terms of the RHS against the LHS (Fig.~\ref{fig8}b) reveals nothing. Indeed, the second term of the RHS (the expectation of trait change) is almost never negative, indicating that mutations on this trait are never deleterious. All observed changes in this trait are consistent with neutral drift. 

\section*{Discussion}
``Information" is a modern buzzword. Some go so far as to claim that ``everything is information"~\citep{Wheeler1990}, but certainly within molecular biology the idea that information is stored in our genes has strong support since the discovery of DNA.
But it certainly has not always been clear how we should measure the amount of information coded in genes, and in particular whether and how selection could act on it. Clearly, information is not the same thing as ``the set of all nucleotides" within the chromosomes of an organism: information must be a ``difference that makes a difference"~\citep{Bateson1972} (or at least, that {\em can} make a difference). Shannon's theory of information~\citep{Shannon1948,CoverThomas2006} provides precisely the framework that is necessary to tell the difference between sequence and information: information is that which allows the bearer to make predictions with accuracy better than chance. Organisms constantly make predictions about their environment (even if this is not always immediately apparent). The metabolic genes, for example, make predictions about the nature and form of resources that the organism is likely to encounter, and developmental genes ``predict" how to make the organism using only the genetic code and the environment it finds itself in. 
A gene in males that encodes a trait that leads to sexual selection predicts that there are females around that are susceptible to this trait. In a sense then, it may be that all traits that have a selective value are informative. But how can we measure this information? We have previously proposed that information can be measured in populations that adapt to an environment by measuring the frequency of substitutions at individual nucleotide positions: those that do not encode information are not under selection (and therefore should be randomly drifting), while those that encode information should be fixed as any mutations would have severe fitness consequences~\citep{Adami1998,AdamiCerf2000,Adamietal2000,Adami2002,Adami2004}. Such a measure was previously applied to DNA binding motifs~\citep{Schneideretal1986} (where the problem of possible higher-order interactions between sites is much reduced) and is used ubiquitously to display sequence motifs~\citep{SchneiderStephens1990}. While this method is most easily used to study how information evolves in computational simulations of evolution~\citep{Schneider2000,HintzeAdami2008} or digital evolution~\citep{Adamietal2000,Ofriaetal2008}, the usefulness of this measure in characterizing the evolving functionality of biomolecules (as opposed to binding sites or digital molecules) has also been shown repeatedly~\citep{Carothersetal2004,Materonetal2004,Carothersetal2006,Carothersatal2006b,Hazenetal2007,GuptaAdami2014}.

It is clear that natural selection (acting on fitness) is responsible for creating the allele frequency differences that are ultimately responsible for the information, and it could be argued that, as a consequence, information and fitness are one and the same thing. They are not, as fitness is measured in offspring per unit time and information is dimensionless and measured in bits (or any other convenient unit). There is no doubt, however, that there is a deep link between the two, and recent work has progressed in unraveling that link~\citep{RivoireLeibler2011}. Here, we decided to test explicitly if information--defined as a measurable (and thus phenotypic) trait--is under selection. In principle, this information is independent of the actual sequence that encodes the information (the same information could be encoded by very different sequences) while in practice we find that the same information is encoded by similar sequences, most likely because inheritance enforces this.  Information cannot be broken down to ``letters": the information content of a single sequence is not defined, unless that sequence is either compared to a specific ensemble~\citep{Schneider1997}, or if an ensemble of sequences is created from that sequence, as we have done here. Of course, each individual trait that contributes to fitness has an underlying genetic encoding, and each such trait has an informational value. The avidians we have used for the experiments we reported here also have individual traits (their capacity to carry out any of the nine different logical functions, or aspects of their gene sequence for self-replication), and the information necessary to carry out each task can be measured individually~\citep{Ofriaetal2008}. But the total information content is a complicated function of the information encoding each individual trait, so it is possible that the total information content is the single best predictor of an organism's success. Indeed, this may ultimately turn out to be the enduring advantage of using an information-based description: if information is the best proxy for fitness, then measuring information is sufficient to infer fitness (which by itself is of course notoriously difficult to measure or estimate, see, e.g.,~\citealt{Orr2009}).

By using information as trait value, we can study the degree to which that trait is under selection using the Price equation. We acknowledge that the value of the Price equation--as wells as its meaning--have been discussed at length in the literature (see, e.g, the review by~\citealt{Frank2012a}). We corroborate here that the Price equation is indeed an identity that ``partitions evolutionary change into two components"~\citep{Frank2012a}, and that what makes the equation useful is that the two components 
play very different roles, and are sensitive to the dynamics. For example, during evolutionary stasis the two terms must cancel each other exactly, whereas during periods of intense positive or negative selection, they take on values significantly different from their stasis values. This is true in the experiments we have performed here too: we can look at the first and second terms of the Price equation averaged over parts of the evolutionary trajectory (a selective sweep or a period of stasis) and notice dramatic changes in the mean values of the first and second term over that period (data not shown).  We also stress that the monotonic increase in information can only hold in a fixed environment. An abrupt change in the environment is almost always going to lead to an abrupt {\em drop} in information, as what used to be information (predicting the previous environment's features) may after the change have turned to entropy. We expect after the change to witness negative covariance of the ``old" information with fitness, and a positive second term, as mutations create new information to replace the old.   

While information content (as a phenotypic trait) has a number of appealing characteristics, we hasten to add that there are numerous caveats associated with it, which (at least today) give fitness measurements the edge. For estimating information content, we cannot rely on the substitution patterns obtained from small populations, as the shared descent of molecules creates a spurious signature of information~\citep{Huangetal2004}. To remedy this, we must generate artificial ensembles by creating all point-wise single mutants, double mutants, and so on, of a sequence. Neglecting the high-order correlations between sites is the only feasible computational shortcut, but this can lead to important biases (mostly under-estimating the information content, see~\citealt{GuptaAdami2014}). However, we believe that thinking in terms of information rather than in terms of fitness can help us navigate the sometimes bewildering jungle of evolutionary genetics, and put evolutionary theory on an even more solid quantitative footing.

\noindent {\bf \sc Acknowledgements}\\
We thank Charles Ofria and Anurag Pakanati for extensive discussions on information estimation, and Thomas LaBar for insightful comments on the manuscript. This work was supported in part by the National Science FoundationÕs BEACON Center for the Study of Evolution in Action, under Contract No. DBI-0939454. We wish to acknowledge the support of the Michigan State University High Performance Computing Center and the Institute for Cyber Enabled Research. 

\bibliography{Myrefs}

\end{document}